\documentclass[twocolumn,10pt]{revtex4}

\usepackage{amssymb}
\usepackage{amsmath}
\usepackage{epsfig}
\usepackage{color}
\usepackage[utf8]{inputenc}
\usepackage{graphicx}
\usepackage{enumerate}
\usepackage{array}
\usepackage{ulem}
\usepackage{soul}
\usepackage{cancel}
\newcolumntype{M}[1]{>{\raggedright}m{#1}}

\newcommand{\be}{\begin{equation}}
\newcommand{\ee}{\end{equation}}
\begin{document}

\title{Looping and Clustering model for the organization of protein-DNA complexes on the bacterial genome}
\author{Jean-Charles Walter}
\affiliation{Laboratoire Charles Coulomb, UMR5221 CNRS-UM, Université de Montpellier, Place Eugène Bataillon, 34095 Montpellier Cedex 5, France}

\author{Nils-Ole Walliser}
\affiliation{Laboratoire Charles Coulomb, UMR5221 CNRS-UM, Université de Montpellier, Place Eugène Bataillon, 34095 Montpellier Cedex 5, France}

\author{Gabriel David}
\affiliation{Laboratoire Charles Coulomb, UMR5221 CNRS-UM, Université de Montpellier, Place Eugène Bataillon, 34095 Montpellier Cedex 5, France}

\author{J\'er\^ome Dorignac}
\affiliation{Laboratoire Charles Coulomb, UMR5221 CNRS-UM, Université de Montpellier, Place Eugène Bataillon, 34095 Montpellier Cedex 5, France}

\author{Fr\'ed\'eric Geniet}
\affiliation{Laboratoire Charles Coulomb, UMR5221 CNRS-UM, Université de Montpellier, Place Eugène Bataillon, 34095 Montpellier Cedex 5, France}

\author{John Palmeri}
\affiliation{Laboratoire Charles Coulomb, UMR5221 CNRS-UM, Université de Montpellier, Place Eugène Bataillon, 34095 Montpellier Cedex 5, France}

\author{Andrea Parmeggiani}
\affiliation{DIMNP, UMR5235 CNRS-UM, Université de Montpellier, Place Eugène Bataillon, 34095 Montpellier Cedex 5, France}

\author{Ned S. Wingreen}
\affiliation{Department of Molecular Biology and Lewis-Sigler Institute for Integrative Genomics, Princeton
University, Princeton NJ 08544, USA}

\author{Chase P. Broedersz}
\email{c.broedersz@lmu.de}
\affiliation{Arnold-Sommerfeld-Center for Theoretical Physics and Center for
  NanoScience, Ludwig-Maximilians-Universit\"at M\"unchen,
   D-80333 M\"unchen, Germany.}

\date{\today}

\begin{abstract}
The bacterial genome is organized by a variety of associated proteins inside a structure called the nucleoid. These proteins can form complexes on DNA that play a central role in various biological processes, including chromosome segregation.
 A prominent example is the large ParB-DNA complex, which forms an essential component of the segregation machinery in many bacteria.
 ChIP-Seq experiments show that ParB proteins localize around centromere-like {\it parS} sites on the DNA to which ParB binds specifically,
 and spreads from there over large sections
 of the chromosome. Recent theoretical and experimental studies suggest that DNA-bound ParB proteins can interact with each other
 to condense into a coherent 3D complex on the DNA.
 However, the structural organization of this protein-DNA complex remains unclear, and a predictive quantitative theory for the distribution
 of ParB proteins on DNA is lacking. Here, we  propose the Looping and Clustering (LC) model, which employs a statistical
 physics approach to describe protein-DNA complexes. The LC model accounts for the extrusion of DNA loops from a cluster of interacting
 DNA-bound proteins that is organized around a single high-affinity binding site. Conceptually, the structure of the protein-DNA complex is determined by a competition between attractive protein
 interactions and the configurational and loop entropy of this protein-DNA cluster. Indeed, we show that the protein interaction strength
 determines the ``tightness" of the loopy protein-DNA complex. Thus, our model provides a theoretical framework to quantitatively compute the binding
 profiles of ParB-like proteins around a cognate ({\it parS}) binding site.
\end{abstract}

\maketitle

\section{Introduction}
Understanding the biophysical principles that govern chromosome structure in both eukaryotic and prokaryotic cells remains an outstanding challenge~\cite{Dekker,Jun,Scolari,Marenduzzo,Dame2011,Emanuel,Mirny}.
 Many bacteria have a single chromosome with a length three orders of magnitude longer than the cell itself, posing a daunting organizational problem.
 Owing to recent technological advances in live-cell imaging and chromosome conformation capture based approaches,
 it is becoming increasingly clear that the DNA  is not coiled like a simple amorphous polymer inside the cell~\cite{Umbarger,Viollier2004,Le2013},
 but rather exhibits a high degree of organization over a broad range of lengthscales~\cite{Lagomarsino}.
 It remains unclear, however, how this spatial and dynamic organization of the chromosome is established and maintained inside living bacteria~\cite{Wang2013}.
 A host of Nucleoid-Associated Proteins (NAPs) have been shown to play a central role in the spatial organization of the bacterial
 chromosome~\cite{Wang2013,Dillon2010,Dame05}. Such NAPs bind to the DNA in large numbers, and by interacting with each other
 and with DNA in both sequence-dependent and sequence-independent manners they can collectively structure the DNA polymer and control chromosome organization.

 In many bacterial species, the broadly conserved ParAB{\it S} system is responsible for chromosome and plasmid segregation \cite{Mohl,Wang2013}. A central component of this system is the partitioning module, which is formed by a large  protein-DNA complex of ParB proteins that assembles around centromere-like {\it parS}
 sites, frequently located near the origin of replication. The ParB{\it S} complexes can subsequently interact with ParA ATPases,
 leading to the segregation of replicated origins \cite{Banigan,Ptacin,Lim,Walter,LeGalletal-1,Vecchiarelli14,Jindal,Surovtsev}. How is this ParB{\it S} partitioning module physically organized on the DNA? ParB is known to bind specifically to {\it parS}, triggering the
formation of a large  protein-DNA cluster, which is visible as a tight focus in microscopy images of fluorescently
 labeled ParB~\cite{Sanchez, Breier,Mohl,LeGalletal-1}. The propensity of ParB to form foci around {\it parS} has been exploited in recent studies,
 which used exogenous expression of fluorescently labeled ParB along with {\it parS} insertion to label DNA loci for live-cell imaging~\cite{Saad,Chen}.
 In the F-plasmid of {\it Escherichia coli} cells, each ParB focus contains roughly $300$ proteins,
 together representing $90\%$ of all ParB present in the cell~\cite{Sanchez}. High-precision ChIP-Seq experiments
on this system provide quantitative ParB binding profiles along the DNA, which are strongly peaked around {\it parS} with a broad decay
 over a distance of up to 13 kilobasepairs (kb), consistent with earlier observations~\cite{Breier,Rodionov}.

\begin{figure}
\center
\includegraphics[width=\linewidth]{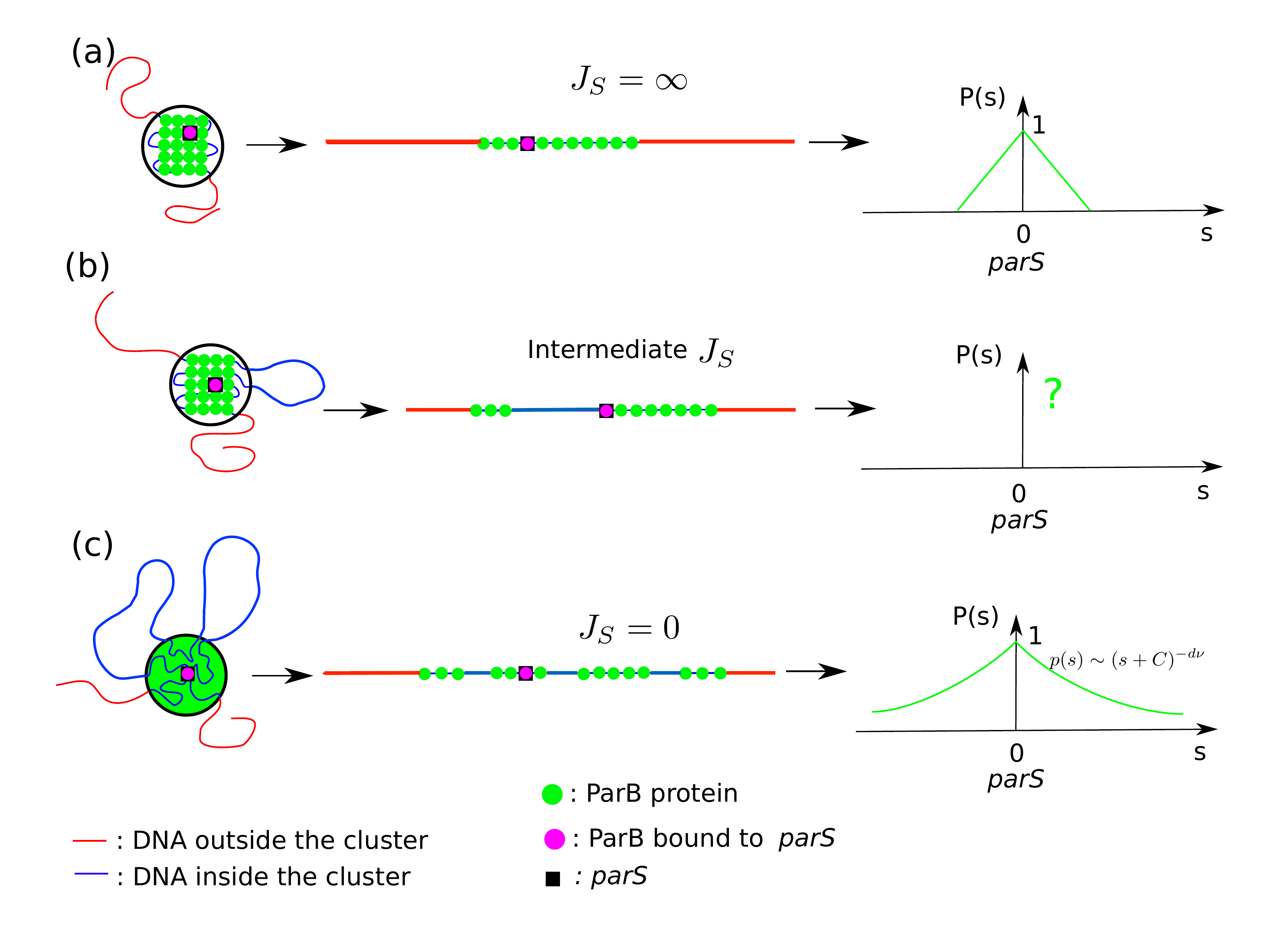}
\caption{\label{fig:schematic}
Schematic illustration of two recent models proposed to describe the
ParB partition complex  (left) accompanied with a typical distribution of ParB
on extended DNA (middle), and
the average distribution profile (right). The Spreading \& Bridging
model~\cite{Broedersz} is shown  with (a) strong coupling $J_S\to \infty$, where thermal fluctuations cannot
break the bonds between proteins such that all bridging and spreading
interactions are satisfied, and (b) intermediate coupling where the
energetic cost of breaking a spreading bond competes with the configurational and loop entropy. With the Looping and
Clustering approach presented here, we propose a simple analytic description for this regime.
(c) The Stochastic Binding model assumes a spherical region of high concentration of ParB
 around {\it parS}~\cite{Sanchez}. This model can be seen as taking the
 limit of the spreading bond strength to zero ($J_S\rightarrow 0$),
  and thus the formation of loops is not hampered by protein-protein bonds. In this limit, the binding
profile can be described as the return of the polymer to an origin of
finite size,  such that the profile is given by
$P(s)\propto(s+C)^{-d \nu}$, where $d$ is the dimension, $\nu$ is the Flory exponent, and $C$ is a constant.}
\end{figure}

 Various models have been introduced to explain the distribution of ParB along DNA around {\it parS} sites.
  An early study of the distribution of ParB proposed that ParB proteins spread from the {\it parS} sequence by nearest-neighbor interactions,
forming a continuous filament-like structure along the DNA~\cite{Rodionov}. This model was termed the Spreading model. However, this is effectively
 a 1D model with short range interactions. On general statistical physical grounds, such a 1D model cannot be expected to account for the formation
 of a large coherent protein-DNA complex, given physiological protein interaction strengths~\cite{Broedersz}. Furthermore, the number of ParB proteins available in the cell is not sufficient to allow enrichment by simple 1D polymerization of ParB along DNA at genomic distances   from {\it parS} as large as observed experimentally~\cite{Sanchez}. To resolve the puzzle of how ParB proteins organize around a {\it parS} site, we recently introduced
 a novel theoretical framework to study the collective behavior of interacting proteins that can bind to a DNA polymer~\cite{Broedersz}.
This model suggested
 that ParB assembles into a three-dimensional complex on the DNA, as illustrated in Figure~\ref{fig:schematic}a,b. Single molecule experiments provided direct evidence for the presence of 3D bridging interactions between two ParB proteins
 on DNA~\cite{Graham,Taylor}. We showed that a combination of such a 3D bridging bond and 1D spreading bonds between ParB proteins constitutes a minimal model for the
 condensation of ParB proteins on DNA into a coherent complex~\cite{Broedersz}, consistent with the observation that ParB-GFP fusion proteins form a tight
 fluorescent focus on the DNA~\cite{Sanchez, Breier,Mohl,LeGalletal-1}.

The statistical properties of the 3D structure of ParB-DNA complexes determines the binding profile of ParB on DNA, which can be accurately measured in ChIP-Seq experiments. However, it is computationally demanding to simulate these binding profiles with the Spreading \& Bridging model. The protein binding profiles  can be easily calculated analytically in the limit of strong protein-protein
 interactions, where the cluster of ParB on the DNA becomes compact with a corresponding triangular distribution of ParB along DNA. The protein binding profiles can also be estimated in the limit of weak protein-protein interactions with the so-called
 Stochastic Binding model, where  a sphere of high ParB concentration  is assumed to exist within which a DNA polymer freely fluctuates~\cite{Sanchez} (see  Figure~\ref{fig:schematic}c).
 The description of the average protein binding profile is thus similar to the return statistics of the polymer into the ParB sphere~\cite{Gennes},
 suggesting a long range (power-law) distribution of ParB proteins along DNA. Importantly, however, neither of these two existing approaches provide a simple way of computing ParB binding profiles around {\it parS} sites over the full relevant range of system parameters. In addition, it remains unclear how the Spreading \& Bridging model and the Stochastic Binding model relate to each other.

Here, we propose a theoretical approach to describe the distribution of ParB proteins around {\it parS} sites on the DNA  in terms of molecular interaction parameters and protein expression levels.
To this end, we develop a simple model for protein-DNA clusters that explicitly accounts for the competition between  the positional entropy associated with placing the loops on the cluster, which favours a looser cluster configuration, and both protein-protein interactions and loop closure entropy, which tend to favour a compact cluster.
 This  {\it Looping and Clustering} model represents a reduced, approximate version of the full Spreading \& Bridging model
 that incorporates the key physical ingredients needed to provide a clearer understanding and at the same time greatly facilitates calculations of the distribution profile of ParB (or other proteins that form protein-DNA clusters). Thus, our approach can be used to estimate molecular interactions between proteins from experimentally determined protein binding profiles.

\section{The Looping and Clustering model} 

To theoretically describe the protein binding profiles of ParB on DNA, we first
consider a DNA polymer of length $L$ that can move in space on a 3D cubic lattice and with a  finite number of proteins $m$. Since the number of ParB proteins in the protein-DNA cluster has been observed to include the vast majority of proteins in the cell~\cite{Sanchez}, we employ a canonical ensemble with a fixed number of ParB proteins $m$ in the ParB complex.
These proteins are able to
diffuse along the DNA. Importantly, in this model the DNA itself is also dynamic and
fluctuates between different three-dimensional configurations, which
are  affected by the presence of interacting DNA bound proteins. When proteins are
bound to the DNA, they are assumed to be able to interact attractively with each other by contact interactions in two distinct ways: (i) 1D
spreading interactions with coupling strength $J_S$, defined as an interaction between proteins on nearest-neighbor sites
along the polymer, and (ii) a 3D bridging interaction with strength
$J_B$ between two proteins bound to sites on non-nearest neighbor-sites on the DNA, but which are positioned at
nearest neighbor-sites in 3D space (see Figure~\ref{fig:schematic}a,b). Thus, these
 bridging interactions couple to the 3D configuration of the DNA, while the 1D spreading interactions do not. Single-molecule experiments provide evidence for bridging bonds~\cite{Graham}, with the
bridging valency of a ParB protein limited to one~\cite{Leonard,Fisher}. Even in this case where each protein can form two spreading bonds and a single bridging bond, the system has been shown to exhibit a condensation transition where the majority of the proteins
form a single large cluster that can be localized by a single {\it parS} site on the DNA~\cite{Broedersz}.

While it is possible to perform Monte Carlo simulations of the Spreading
\& Bridging model for a lattice polymer, such simulations are computationally demanding. In this paper, we aim to provide a simple analytical description for the average binding profile of proteins along the DNA
  (see right panels in Figure~\ref{fig:schematic}). With this aim in mind, we
can  simplify our description by realizing that the
configurations of ParB proteins
  along the DNA are  more sensitive to $J_S$ than to $J_B$, for sufficiently large $J_B$.
While both spreading and bridging bonds are necessary for the
condensation of all proteins into a single cluster,
  loop extrusion from the cluster is controlled by $J_S$, and such loop
extrusion  strongly impacts the binding profile of proteins on the DNA. Indeed, a loop can be extruded from the protein-DNA cluster by breaking a spreading bond, but without effecting the internal configuration of the bridging bonds.  Therefore, we will assume that $J_B$ is sufficiently large to maintain  a coherent 3D protein-DNA cluster, leaving $J_S$ as the main adjustable parameter in the model.

  A contiguous 3D cluster of proteins on DNA with loops can effectively
be represented graphically by a disconnected 1D cluster along the DNA, where
connections in 3D between the  1D subclusters are implied, and
domains of protein-free DNA
within the disconnected 1D cluster represent loops that emanate from the
3D cluster (see Figure~\ref{fig:schematic}b,c). We can describe this system by a  reduced model for the
  effective 1D cluster in which we  account for the entropy of the loops
that originate from the protein-DNA cluster.
  In this model, the spreading bond energy set by the parameter $J_S$ combined with the cost in loop closure entropy,
competes with the positional entropy for placing loops on the cluster and will therefore play a crucial
role in  determining the binding profile of ParB on DNA around a {\it parS} site.

To capture these effects, we propose the reduced Looping and Clustering
(LC) model, which offers a simplified description of 3D protein-DNA clusters with spreading and bridging bonds.
In this model a loop is formed whenever there is a gap between 1D clusters. We can make the connection between the gaps
in the 1D
  cluster and the number of loops extending from the 3D cluster explicit
by writing down the partition function for this model.
  The effective 1D cluster corresponding to a 3D cluster with $m$
proteins and $n$ loops has a multiplicity:
\begin{equation}
\Omega_{\rm cluster}= \frac{(m-1)!}{(m-n-1)!n!},
\end{equation}
which counts the number of ways in which one can partition $m$ proteins
into $n+1$ subclusters in 1D. This multiplicity leads to a positional entropy of mixing, $S_{\rm cluster} = \ln \Omega_{\rm cluster}$, for placing  $n$ loops at $m-1$ possible positions (in units of $k_{\rm B}$). Note, we do not explicitly include the number of ways in which the bridging bonds can be formed, since loop formation is not expected to substantially affect the possible configurations of bridging bonds. However, creating $n$ loops will
require breaking  $n$ spreading bonds, and the probability at equilibrium for this
  to occur will include a Boltzmann factor  $\sim \exp\left(-n J_S\right)$, where the interaction energy is expressed in units of $k_BT$. Within our simple description, we do not consider how the formation of a loop affects the full internal entropy of the protein-DNA cluster, but this can be expected to be a fixed number per loop that can be absorbed into $J_{\rm S}$.
Furthermore, the loops that are formed are assumed to be independent, and thus contribute to the loop closure entropy (in units of $k_{\rm B}$) as~\cite{Gennes}:
\begin{equation}
S_{\rm loop } =-d\nu\sum_{i=1}^{n} \ln(\ell_i)\,,\label{entropy}
\end{equation}
where $d$ is the spatial dimension, $\nu$ is the Flory exponent, and the loop length is measured in units of the lattice spacing of the polymer $a$,
 which we take to be equal to the footprint of a ParB protein, e.g. $16~{\rm bp}$ for the exogenous ParAB{\it S} system of {\it E. coli} \cite{Sanchez}.

This entropy is obtained by considering both the loops formed within the protein cluster and the protein-free segment of DNA outside the cluster.
Indeed, the number of configurations associated with loop $i$ for a Gaussian polymer is given by $z^{\ell_i}\ell_i^{-d\nu}$~\cite{Gennes,Hanke},
 where $z$ is the lattice coordination number. Therefore, there is also an extensive contribution to the entropy given by $k_{\rm B} \ell_i \log(z)$.
 However, when a loop of length $\ell_i$ forms, the same length of polymer is removed from the DNA outside of the cluster, which also results in
 a reduction of the entropy by $k_{\rm B}\ell_i \log (z)$. Thus, there is a precise cancelation between the extensive contribution to the entropy
 associated with the loop inside the cluster and the extensive contribution due to effectively shortening the DNA outside the cluster~\endnote{Although this reasoning is not strictly true for self-avoiding polymers, it does hold if we adopt the usual approximation used in the Poland-Scheraga model for DNA melting that self-avoidance acts only within individual loops}.

It is now straightforward to write down the partition function of the Looping and Clustering model:
\begin{widetext}
\begin{eqnarray}
  \label{eq:Z}
Z_{LC}&=& \sum_{n=0}^{m-1} \frac{(m-1)!}{(m-n-1)!n!}\exp(-n J_S)
\int_{\ell_0}^{\ell_{\rm max}} d \ell_1 \ell_1^{-d \nu}
...\int_{\ell_0}^{\ell_{\rm max}}d \ell_n \ell_n^{-d \nu}\nonumber \\
&\underset{\ell_{\rm max} \rightarrow \infty}{\longrightarrow}&
\left[1+\exp(-J'_S)\right]^{m-1}.\label{PF}
\end{eqnarray}
\end{widetext}
where 
$J'_S = J_S + \ln \left[ \ell_0^{d \nu - 1} (d \nu - 1) \right] > J_S$  is a renormalized loop activation energy that includes the cost in loop closure entropy). All lengths are measured in units of the protein's footprint $a$, $\ell_0$ is the lower cutoff of loop sizes and approximately
 represents the persistence length of DNA, and the bond interactions are in units of $k_{\rm B} T$. In the partition function,
 we conveniently set the upper boundary of integration, $\ell_{\rm max}$, to infinity. Strictly speaking, the upper boundary for $\ell_j$ should be
$L-(m+L_j)$, where $L_j=\sum_{i=1}^{j-1} $ represents the total
accumulated loop length before loop $j$. In practice, however, for chromosomes, but arguably also for plasmids, $L\gg m$
and the probability to have a large loop is very small. For instance, if we consider the F-plasmid of {\it E. coli} with a length of 60 kbp,
we have $L=3750$ in units of the ParB footprint of 16 bp~\cite{Bouet09,Sanchez}.
 For this system, Monte Carlo (MC) simulations (see Appendix A) of the LC model, with $m = 100$ reveal that the average
cumulated loop size is $\approx 500$ for small couplings ($J_S = 1$) down to $\approx25$ for large couplings ($J_S = 4$), which in
both cases is much less than the DNA length. Thus, for biologically relevant cases it is reasonable to assume that the
length of the DNA polymer is much larger than the footprint of the whole protein complex on the DNA.

The LC model constitutes a simple statistical mechanics
approach to describe how proteins assemble into a protein-DNA cluster
with multiple loops. Next, we will include a {\it parS} site on the DNA, to which ParB proteins bind with a higher affinity than the other non-specific binding sites on the DNA. Our central aim is to compute the binding profile of ParB around this {\it parS} site.

\section{Profile of Par B for fixed number and sizes of loops} 

With our approach, we aim to quantitatively describe average ParB binding profiles, which are directly measurable by ChIP-Seq experiments. By fitting our model to such  ChIP-Seq data, it would be possible to extract microscopic parameters such as the number of proteins in the ParB clusters and the protein-protein interaction parameters such as $J_S$. In this section, we will describe how to compute the ParB binding profile around this {\it parS} site given a fixed number of loops with specified loop lengths. Then we will use the statistical mechanics framework provided above, to
perform a weighted average over all possible loop numbers and sizes to arrive at a simple predictive theory for the ParB binding
profile.

\subsection{1-loop binding profile} 

\begin{figure}
\center
\includegraphics[width=1\linewidth]{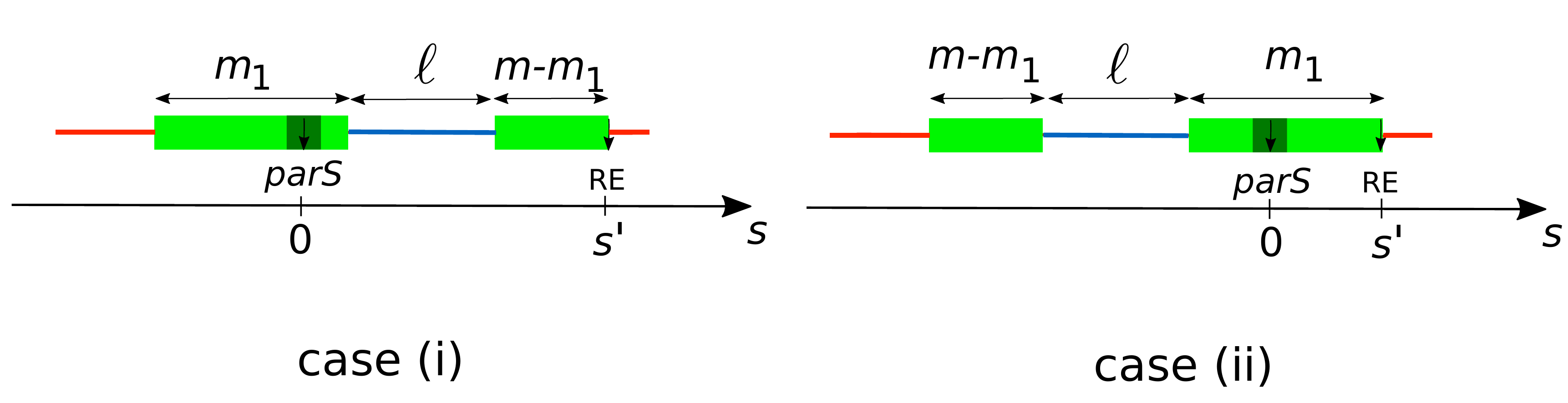}
\caption{\label{scheme} Schematic of the system with $m$ proteins and a single loop of size $\ell$. The whole
cluster is split in two parts: $m_1$ is the number of proteins in the cluster that overlaps with {\it parS} and $m-m_1$ is
the number of proteins in the other cluster.
The origin of the genomic coordinates is {\it parS}, the right edge of the system (RE) is located at the coordinate $s'$.
We can divide the configurations into two equally likely cases:
(i) the leftmost cluster overlaps with {\it parS} or (ii) the rightmost cluster overlaps with {\it parS}.
}
\end{figure}

It is instructive to start our analysis of ParB binding profiles by first calculating the probability of ParB occupancy as a
 function of distance from the {\it parS} site for
the case of a protein-DNA cluster with only one DNA loop ($n=1$) with fixed loop length $\ell$. We will assume a fixed
number $m$ of ParB proteins in this 1-loop protein-DNA cluster, and that one of these proteins is bound to the {\it  parS} site at any time,
 as illustrated in Figure~\ref{scheme}. Thus, to calculate the 1-loop ParB binding probability, $P_1(s, \ell)$, at a distance $s$
 from {\it parS}, we need to consider all possible configurations of proteins in the protein-DNA cluster subject to these constraints.

First, we note that $P_1(s, \ell)=0$ for $s>m+\ell$, because the 1D cluster can maximally extend to a distance $m+\ell$,
 which occurs when the 1D cluster adopts a configuration that lies entirely on one side of the {\it parS} site.
 For a binding site at a distance $s<m+\ell$, the ParB binding probability is  reduced, either by configurations where this site is
 located on the DNA loop within the 1D cluster, or by states where the 1D cluster adopts a configuration around the {\it parS} site
 that does not extend to the binding site at $s$, placing this site outside the 1D cluster. To capture these effects, it is helpful
 to express $P_1(s, \ell)$ in terms of conditional probabilities:
\begin{eqnarray}
 P_1(s,\ell)&=&P_1(s,\ell | \text{loop$(s)$})p_{{\rm loop}}(s)+P_1(s,\ell | \overline{\text{loop$(s)$}})p_{\overline{{\rm loop}}}(s)\nonumber\\
&=&\ P_1(s,\ell | \overline{\text{loop$(s)$}})p_{\overline{{\rm loop}}}(s),
\label{0}
\end{eqnarray}
where ``loop$(s)$'' represents a condition with probability $p_{{\rm loop}}(s)$ corresponding to site $s$ being part of a loop extruding from the cluster,
 i.e. an unoccupied site on the DNA within the protein cluster, as depicted in Figure~\ref{scheme}.
 The overbar here represents the complementary condition, and the expression above simplifies
 because $P_1(s,\ell | \text{loop$(s)$})=0$ by construction.

\begin{figure}
\center
\includegraphics[width= 0.9 \linewidth]{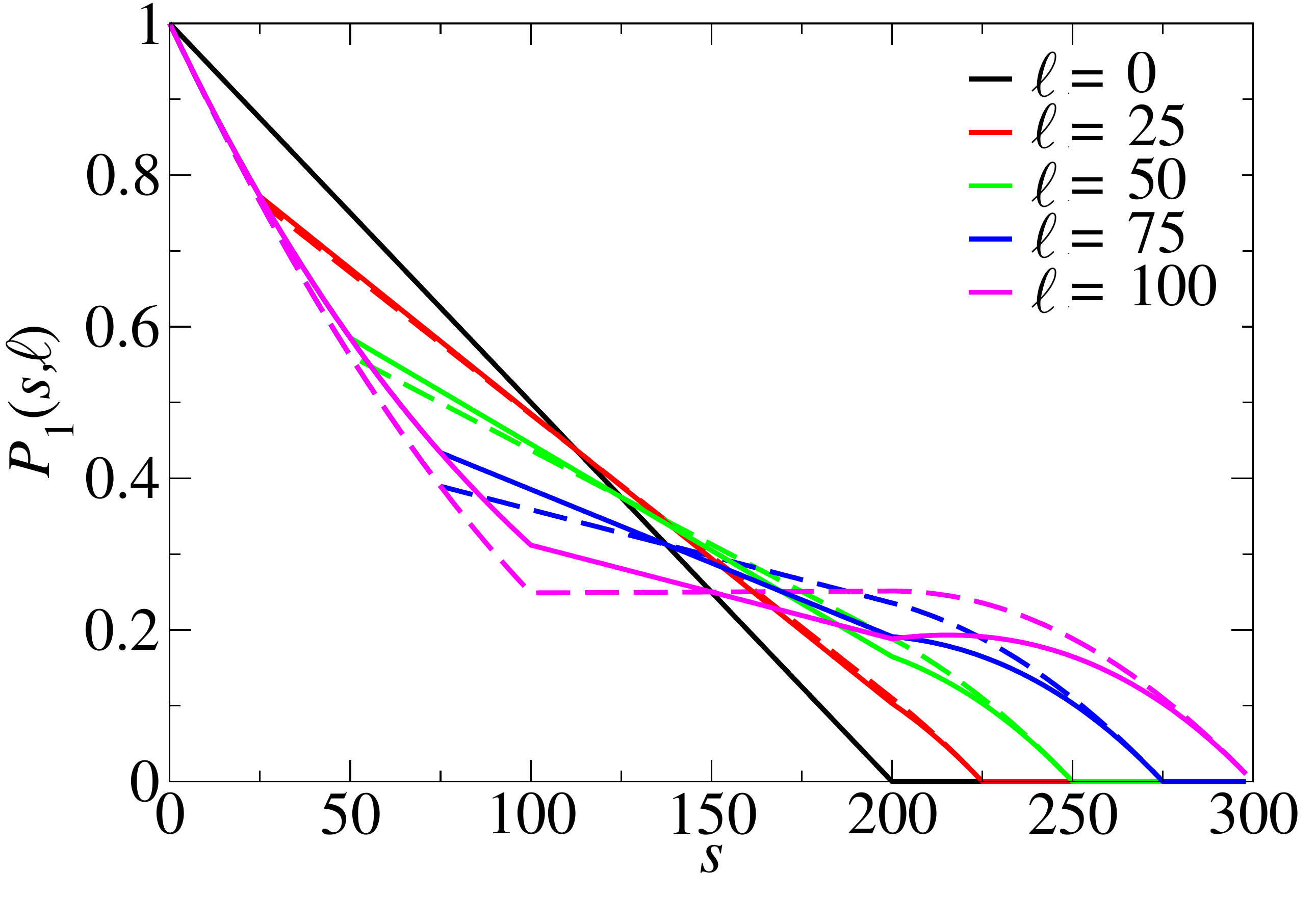}
\caption{\label{fig1}
Protein occupation probability, $P_1(s,\ell)$, for a site a genomic distance $s$ from the {\it parS} site
 for different loop lengths $\ell$ and a fixed cluster size of $m=200$ proteins. Solid curves represent analytic calculations
 from Eqs.~(\ref{0}), (\ref{2}), and (\ref{4}), and dashed curves represent data obtained from exact numerical enumeration for comparison to our analytical approximations.
  We note that for $\ell=0$, we recover the triangular profile of the S\&B model in the strong coupling limit $J_S\to \infty$~\cite{Broedersz}.}
\end{figure}

We can proceed to calculate the conditional probability, $P_1(s,\ell | \overline{\text{loop$(s)$}})$, by decomposing this contributions as a sum
 of probabilities of mutually exclusive configurations, which are conditioned by the location $s'$ of the right edge of the
 1D ParB cluster denoted as ``${\rm end}(s')$" (see Figure~\ref{scheme}).
 Then, we will take a continuous limit for the binding profile assuming $m\gg1$,
 and express the binding profile $P_1(s,\ell | \overline{\text{loop$(s)$}})$ in terms of probabilities, $p_{{\rm end}}(s')$,  for the condition
 describing the position of the right edge of the cluster. Thus, we first write the conditional probability $P_1(s,\ell | \overline{\text{loop$(s)$}})$
 for $s\ge0$ (the case $s<0$ is obtained by symmetry) as:
\begin{eqnarray}
P_1(s,\ell | \overline{\text{loop$(s)$}})&=&\sum_{s'=0}^{m+\ell} P_1(s,\ell | \overline{\text{loop$(s)$}},\text{end$(s')$}) p_{{\rm end}}(s'),\nonumber\\
& \approx& \int_{0}^{m+\ell}ds'\Theta(s'-s) p_{{\rm end}}(s').\label{1b}
\end{eqnarray}
Clearly, $P_1(s,\ell | \overline{\text{loop$(s)$}},\text{end$(s')$})=1$ when $s<s'$ and zero otherwise, and thus we have replaced this term by the Heaviside step
function $\Theta(s'-s)$ and approximated the sum by an integral in the second line above.

To calculate $p_{{\rm end}}(s')$,
it is convenient to introduce two subclusters, $1$ and $2$, with $m_1$  and $m-m_1$ proteins respectively ($0<m_1<m$), such that cluster $1$
with $m_1$ proteins is overlapping with {\it parS}, as shown in Figure~\ref{scheme}.
Given two such subclusters, two equally likely situations can occur:
 (i) the leftmost cluster overlaps with {\it parS}, i.e. $m-m_1+\ell\le s'<m+\ell$ or
 (ii) the rightmost cluster overlaps with {\it parS}, i.e. $0\le s'<m_1$.
 This directly allows us to construct the conditional probability to find the right edge of the whole system, such that one of the $m_1$ proteins
in the cluster overlaps with {\it parS}:
 \begin{eqnarray}
p_{\rm end}(s' | m_1)=&\frac{1}{2m_1}&\left[\Theta(s'-(m-m_1+\ell))\Theta(m+\ell-s') \right. \nonumber \\
&+& \left. \Theta(m_1-s')\right],
\end{eqnarray}
where the prefactor 1/2 comes from the equal probabilities to find the system in one of the two cases (i) and (ii). The conditions (i) and (ii) are encoded with
a product of two unit step functions for (i) and a single step function for (ii). Each single realization can
be obtained by shifting the position of the site in cluster 1 overlapping with {\it parS} and is equally likely,
 giving rise to an overall prefactor $1/m_1$. From this, we can obtain the full probability $p_{{\rm end}}(s')$ by integrating over $m_1$:
 \begin{widetext}
 \begin{eqnarray}
p_{{\rm end}}(s') &\approx&\int_1^{m-1}\,dm_1\,p_{\rm end}(s' | m_1) p(m_1)\nonumber\\
&=&\frac{\Theta(m+\ell-s')}{m(m-2)}\left[\Theta(1-(m+\ell-s'))(m-2)+\Theta(m+\ell-s'-1)\Theta(s'-\ell-1)(s'-\ell-1)\right.\nonumber \\
&+&\left.\Theta(1-s')(m-2)+\Theta(s'-1)\Theta(m-1-s')(m-1-s)\right]
\end{eqnarray}
 \end{widetext}

 where we used $p(m_1)=2 m_1/(m(m-2))$, since the number of configurations to place cluster 1 is $\propto m_1$ and $m_1\in [1,m-1]$.
 Using this expression for the normalized probability distribution for the right edge of the 1D cluster
to be positioned at $s'$,  we can compute the conditional probability in Eq.~(\ref{1b}):
 \begin{widetext}
 \begin{eqnarray}
P_1(s,\ell | \overline{\text{loop$(s)$}})&\approx&\frac{\Theta (\ell+m-s)}{(m-2) m} \left[\frac{(m-2)^2}{2}  \Theta (\ell-s+1)+(m-2) \Theta (\ell+m-s-1)\right.\nonumber \\
&+&(m-2) (m+\ell-s) \Theta(s-m-\ell+1)+(m+\ell-s-1) m-\ell+s-3) \Theta (s-\ell-1)\nonumber\\
&+&\left.\Theta (1-s) \left(\frac{(m-2)^2}{2}  \Theta (m-s-1)+(m-2) (1-s)\right)+\frac{(s-m+1)^2}{2}  \Theta (s-1) \Theta (m-s-1)\right]
\label{2}
\end{eqnarray}
 \end{widetext}

To obtain the full 1-loop protein distribution (Eq.~\eqref{0}), we first need to compute the probability for a site to not be part of loop,
\begin{equation}
p_{\overline{{\rm loop}}}(s)=1-p_{\rm loop}(s).
\end{equation}
If the loop density, $\rho$, were uniform, we would simply have
$p_{\text{loop}}^{\text{uni}}(s)=\ell \rho^{\rm uni}(m,\ell)=\frac{\ell}{m+\ell}$, since the 1D cluster has a total length of $m+\ell$ with a single loop
 of length $\ell$. This uniform condition would only apply if we randomly choose $\ell$ sites to be part of the loop and ignore the requirement that
 all these loop sites need to be neighboring. In a real cluster, however, we expect the loop density $\rho_{\rm loop}(s)$ to be higher in the bulk
of the 1D cluster than close to the {\it parS} site or the edges, because  fewer loops can be formed near the {\it parS} site or near the boundaries
 of the 1D cluster, at which
 a protein must be bound by construction. In particular, we expect the loop density,
 $\rho(s,m,\ell)\propto \min(s,\ell,m+\ell-s)$, which measures the number of ways a site at $s$ can be part of a loop.
 This results in the normalized probability:
\begin{eqnarray}
\label{eq:4}
p_{\rm loop}(s)
&=&\ell\rho(s,m,\ell) \\
&=&\frac{\ell\min(s,\ell,m+\ell-s)}{\Theta(m-\ell)\left[\ell^2+\ell(m-\ell)\right]+\Theta(\ell-m)\left(\frac{m+\ell}{2}\right)^2}\ \nonumber 
\end{eqnarray}
\begin{figure}[h!]
\centering\includegraphics[width=.8\linewidth]{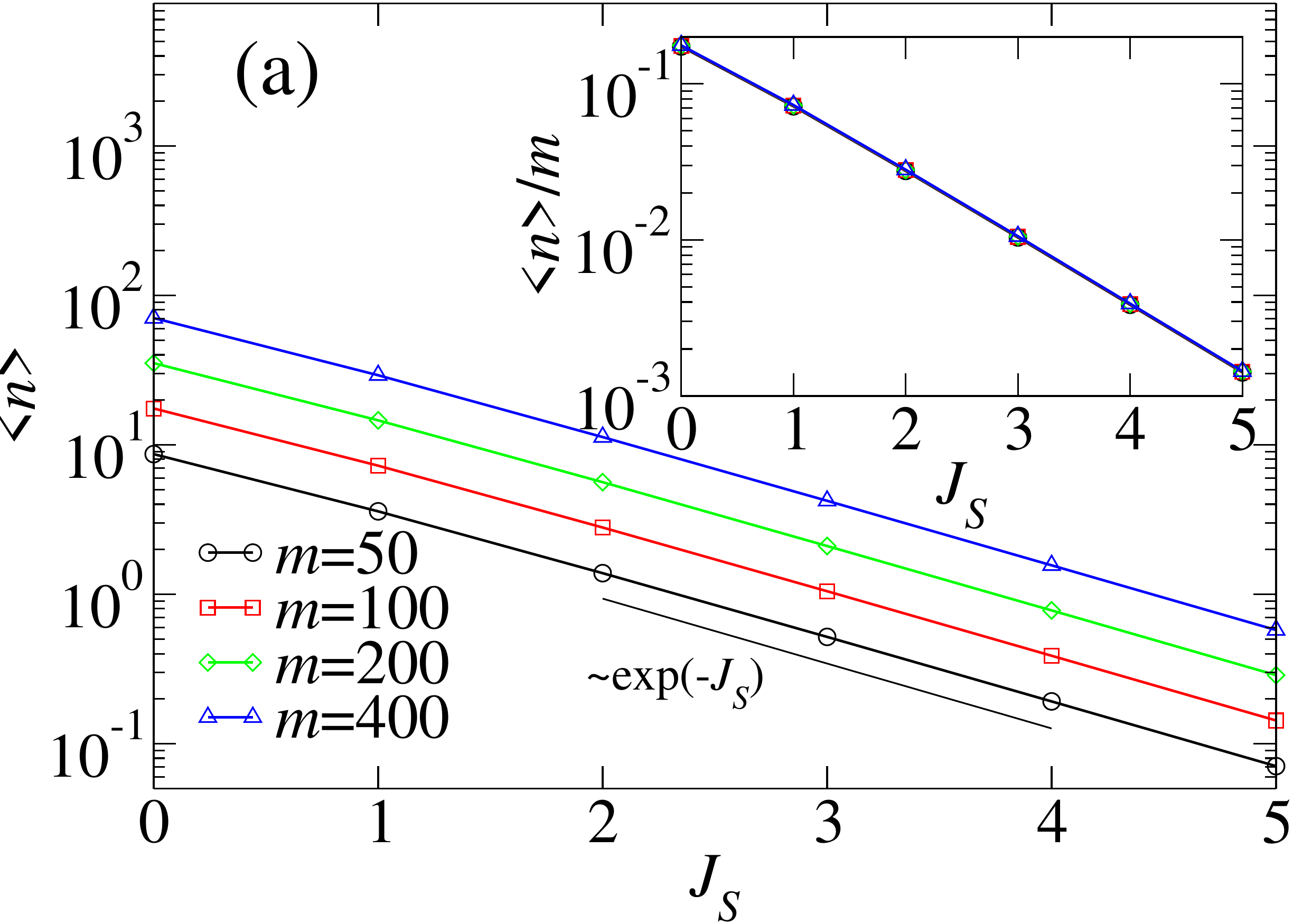}\\
\centering\includegraphics[width=.8\linewidth]{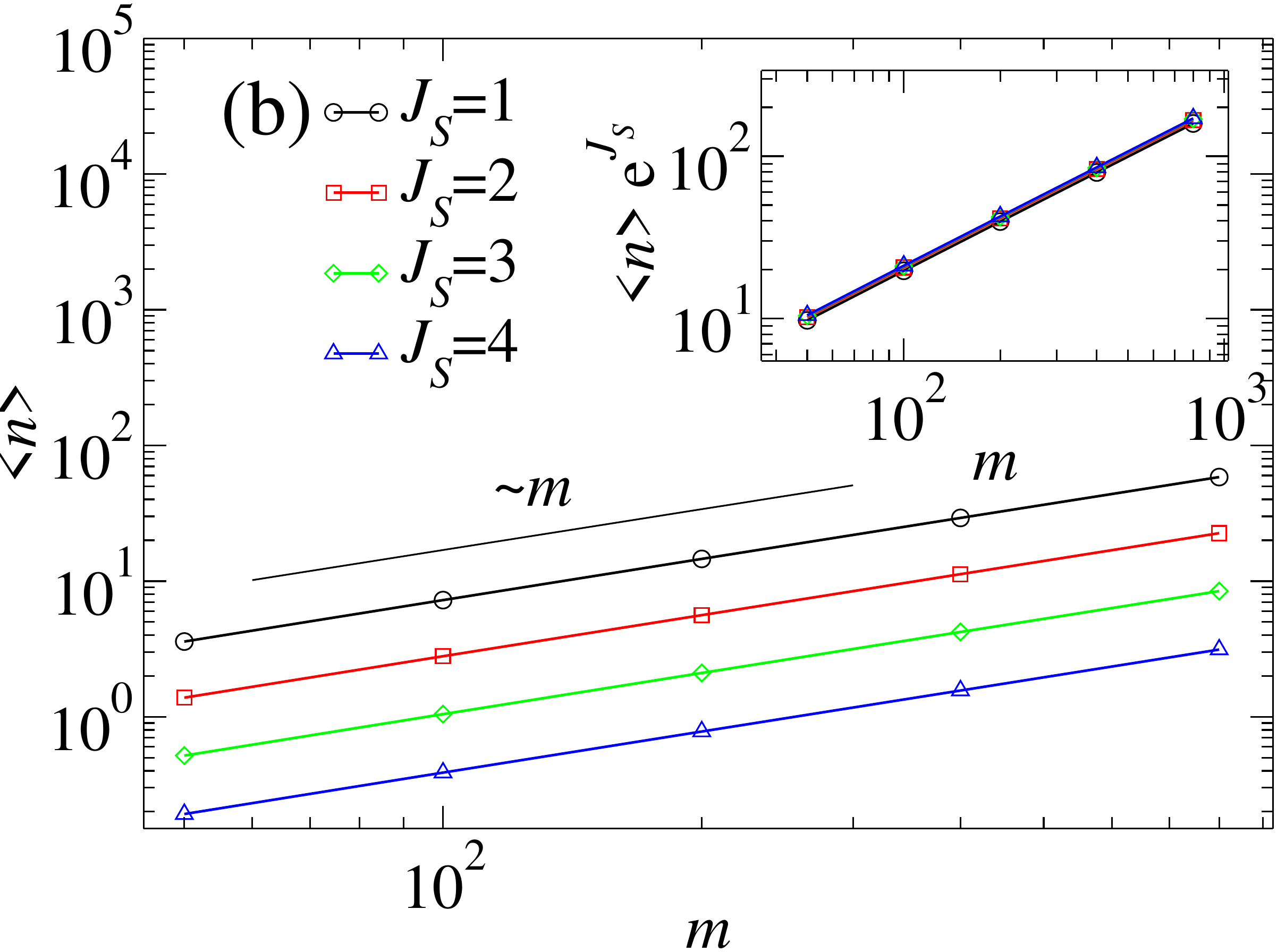}\\
\centering\includegraphics[width=.8\linewidth]{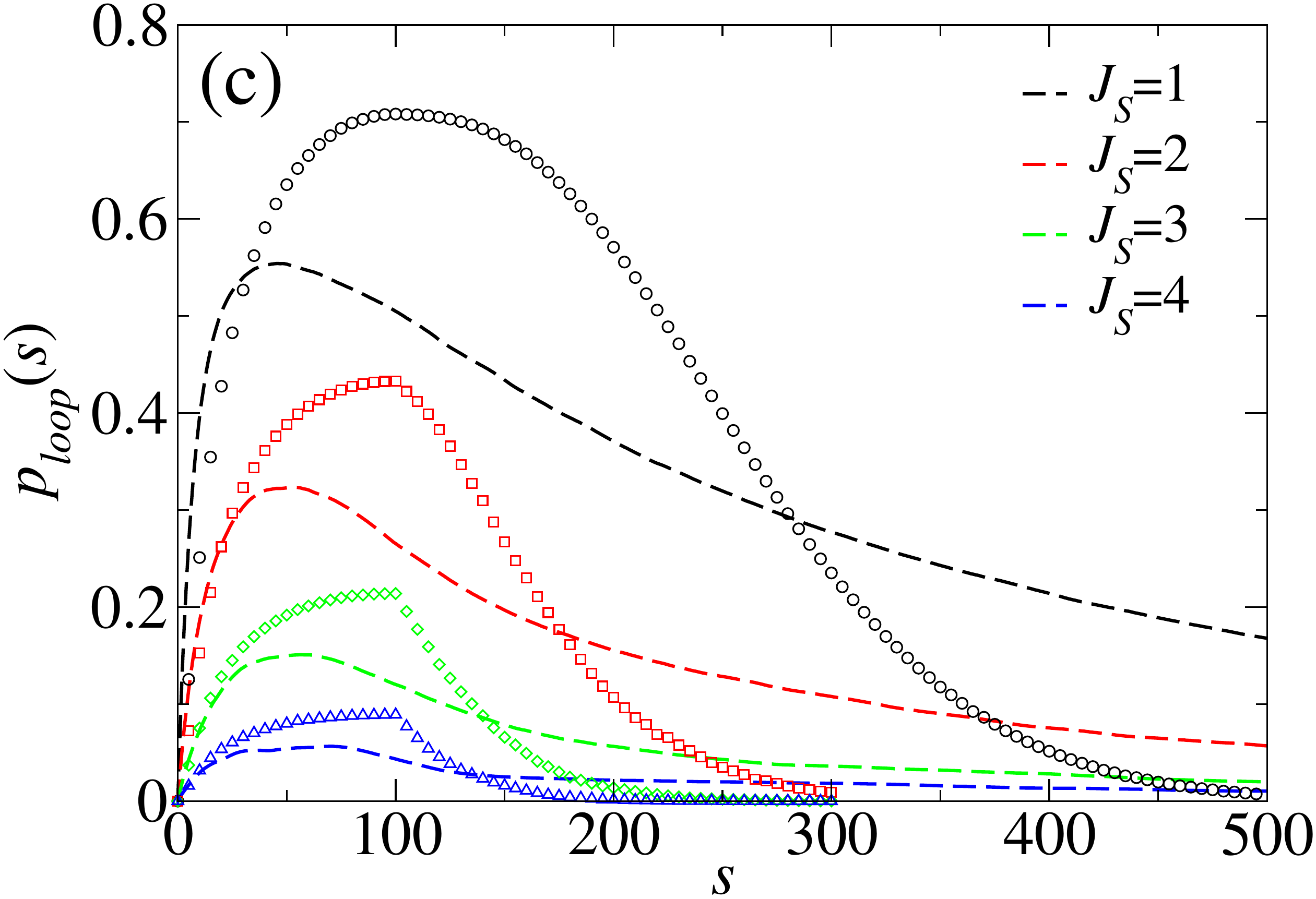}
\caption{\label{fig4}
 (a) Average number of loops, $\langle n\rangle$, as a function of spreading coupling strength $J_S$ obtained from Eq.~(\ref{n1}). The different curves correspond to protein number $m=50$ (black) $m=100$ (red), $m=200$ (green), and $m=400$ (blue),
 with loop-size cutoff  $\ell_0=10$. We observe an exponential decrease $\langle n\rangle\propto e^{-J_S}$ in accordance with Eq.(\ref{n1}).
 Inset: Same data replotted with expected dependence of average loop number on $m$ scaled out.
 (b) Average number of loops $\langle n\rangle$ as a function of $m$ for $J_S=1$, 2, 3, and 4.
The behaviour is linear as expected from Eq.(\ref{n1}).
The prefactor that determines the vertical shift between the different curves scales with $e^{-J_S}$, as demonstrated in the inset of panel (b).
(c) Average loop probability as a function of the genomic coordinate with $m=200$ and $L=4000$ for protein-DNA clusters with fluctuating loop number and loop lengths.  Different curves correspond to different spreading couplings $J_S=1$, 2, 3, and 4.
The  analytic approximation using Eq.(\ref{loopdensity}) for the loop density,  averaged over different loop configurations with the appropriate Boltzmann factor as in Eq.~\eqref{Ptot}  is compared to MC simulations  (dashed curves) of the LC model (see Appendix A).}
\end{figure}

In the normalization of this expression we distinguish the cases where the loop is either smaller or larger than the number of proteins in the cluster.
With Eqs.~(\ref{2}) and (\ref{eq:4}), we have all the elements to calculate the 1-loop protein binding profile $P_1(s,\ell)$ from Eq.~(\ref{0}).

We investigated the binding profiles $P_1(s,\ell)$ predicted by this model for a selected set of parameters, as shown in Figure~\ref{fig1}. We only show $s>0$
 because of the symmetry of the binding profile. It is instructive to contrast these profiles with the triangular profile (black curve) for a cluster with no loops.
 As expected,
the addition of loops widens the profile, allowing the tail of the distribution to extend out to a distance $m+\ell$. The widening of the binding profile is accompanied
 by a faster decay of the profile in the vicinity of {\it parS}, which crosses over to a flatter profile at distances $s>\ell$ due to
 additional contributions from configurations where the loop lies between the $parS$ site and site $s$.

 Interestingly, for large loop size the profile can even become non-monotonic with a slight increase near the far edges of the domain.
 These features of the profile reflect the reduced loop density near {\it parS} and near the far edges of the cluster. Note that the integral under this curve remains constant for varying $\ell$ to conserve the number of particles in the cluster. To verify the validity of the analytical approximations leading to $P_1(s,\ell)$, we used exact enumeration as a benchmark.
 Overall, the numerics (dashed lines) and the analytics (solid lines) are in good agreement for the 1-loop case, as shown
 in Figure~\ref{fig1}.
In the next section, we employ the approximate analytical expressions obtained above, to efficiently calculate the full binding profile averaged over all configurations.

\section{Protein binding profiles and statistics of the Looping and Clustering model}

Above we  defined the Looping and Clustering model and calculated the binding profile of proteins around a {\it parS} site for a cluster
 with 1 loop with fixed length.  Real protein-DNA clusters, however, are expected to fluctuate with new loops forming
 and disappearing continuously. To capture such fluctuations,  we will use the  expressions for the binding profile of a static cluster with
 fixed loop length together with a statistical mechanics description of the LC model to obtain {\it average} binding profiles
 for dynamic clusters, including an ensemble average over both the number of loops and the loop lengths.

To obtain a full binding profile averaged over all realizations, it is useful to investigate the statistics of loops that extend
 from the protein-DNA cluster and how these statistics are determined by the underlying microscopic parameters of the model.
 We start by considering the number of loops that extend from the cluster.
Using the partition function in Eq.~(\ref{eq:Z}), it is possible to calculate the basic features of the LC model.
For instance, the moments of the distribution of the number of loops are given by:
\begin{equation}
\langle n^\alpha\rangle=\frac{1}{Z_{LC}}\sum_{n=0}^{m-1} n^{\alpha}\frac{(m-1)!}{(m-n-1)!n!}e^{-n J_S}\left[\frac{\ell_0^{1-d\nu}}{d\nu-1}\right]^n.\label{n0}
\end{equation}
From this, we find the the average loop number is given by:
\begin{equation}
\langle n\rangle=(m-1)\frac{1}{1+e^{J'_S}}\underset{m, J'_S\gg1} \sim m e^{-J_S},\label{n1}
\end{equation}
where $J'_S$ is the renormalized loop activation energy introduced in Eq.~\eqref{PF}.
 The average loop number $\langle n\rangle$ is depicted in Figure~\ref{fig4}a, demonstrating the exponential dependence on the spreading energy $J_S$.
 In Figure~\ref{fig4}b, we plot  $\langle n\rangle$ as a function of the total number of proteins $m$ in the protein-DNA cluster. Indeed, we observe the expected linear dependence of the average loop number $\langle n\rangle$ on $m$ over a broad range of parameters. These results illustrate how the average number of loops is determined by the competition  between
the effective renormalized loop activation energy, $J'_S$ (including the cost in loop closure entropy), and the gain in the positional entropy of mixing (see Appendix B).

\begin{figure}[h!]
\centering\includegraphics[width=.8\linewidth]{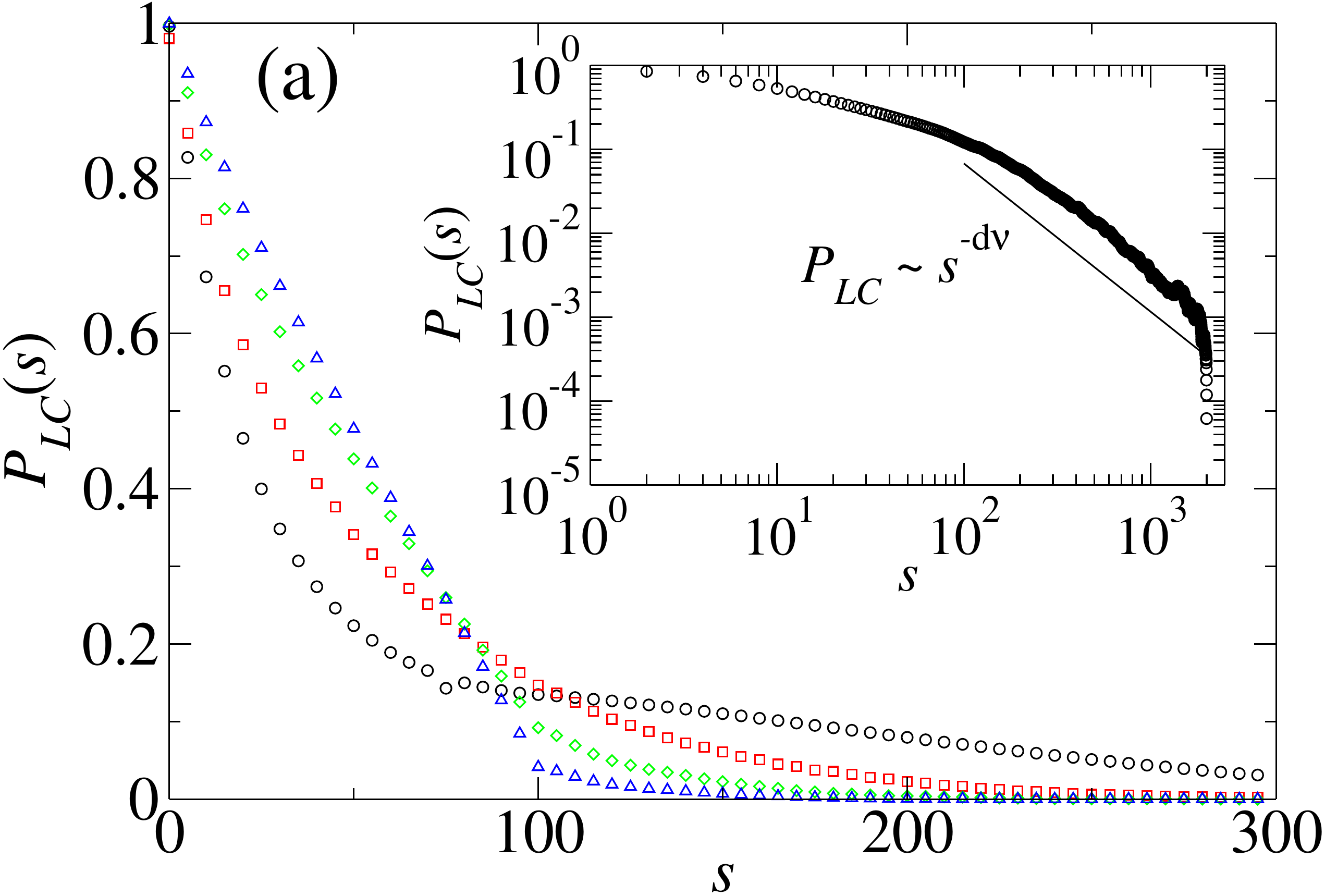}\\
\centering\includegraphics[width=.8\linewidth]{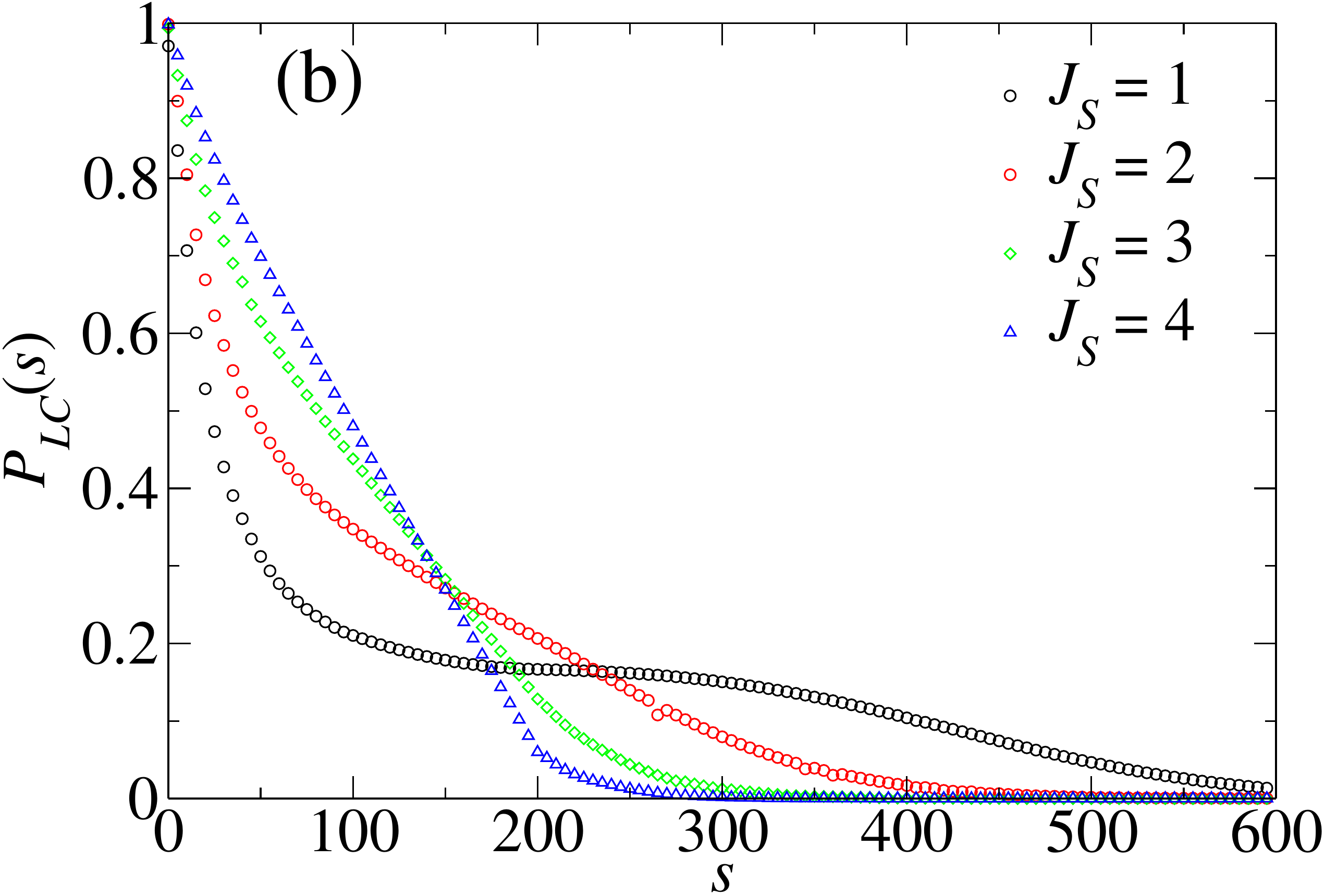}\\
\centering\includegraphics[width=.8\linewidth]{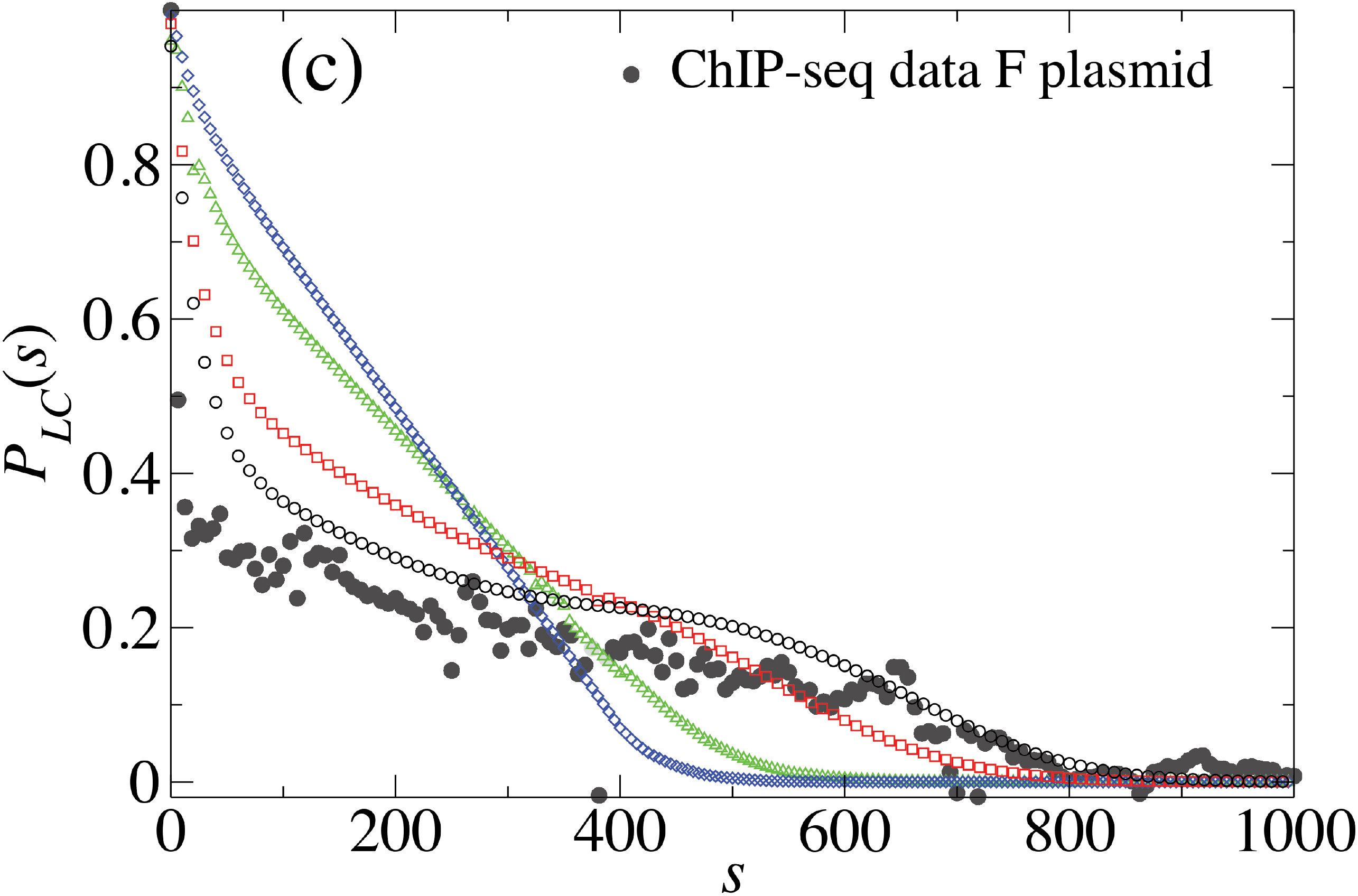}
\caption{\label{fig6}
Binding profiles of ParB from Eq.~\eqref{Ptot} plotted versus the
genomic distance $s$ to  {\it parS} for (a) $m=100$, (b) 200, and (c) 400.
   In Eq.~\eqref{Ptot}, the loop size integrals were calculated with a
lower cutoff $\ell_0=10$ and an upper cutoff of $10 \ell_0$; summations
were truncated at $n=15$. The dark grey circles in panel (c) show
experimental ChIP-Seq ParB enrichment data from the F-plasmid of
 {\it E. coli} extracted from~\cite{Sanchez}. The inset in panel (a) shows the
binding profile of ParB  versus genomic distance $s$ to {\it parS} for
$J_S=1$, $\nu=0.588$ (self-avoiding polymer). The results in this inset
 were obtained by Monte Carlo simulations of the LC model (see SI for details).
The data are plotted in log-log scale, we observe the power law decay
 $P_{LC}\sim s^{-d\nu}$ as expected in the limit of low $J_S$, where
 the LC model becomes conceptually similar to the Stochastic Binding model.}
\end{figure}

The linear dependence on $m$ in Eq.~\eqref{n1} reflects that loops are assumed to be able to form anywhere in the cluster in the Looping and Clustering model.
 However, one would naively expect that loops can
 only form at the surface of a 3D cluster, resulting in a dependence $\langle n\rangle \sim m^{2/3}$ for a  compact, spherical cluster.
 However, Monte Carlo simulations of the full S\&B model have revealed that the protein-DNA clusters are not compact~\cite{Broedersz}, but rather
 have a surface that scales almost linearly in $m$, close to the behavior of the simplified LC model presented here. The non-compact nature of the protein-DNA cluster is perhaps not surprising because each protein can form only one bridging bond.

A closely related statistic is the average accumulated loop length $\langle\ell\rangle$. From the LC partition function, we notice that the loop length is
completely decoupled from the coupling constant $J_S$ and depends only on the upper cutoff $\ell_{\rm max}$. Therefore, the cumulated average loop length becomes:
\begin{equation}
 \langle\ell\rangle \underset{\ell_{\rm max} \gg \ell_0}= \frac{d\nu-1}{2-d\nu}\frac{\ell_{max}^{2-d\nu}}{\ell_0^{1-d\nu}}  \langle n\rangle,
 \label{looplength}
\end{equation}
where the factor in front of $\langle n\rangle$ represents the average length per loop. This prefactor induces a small algebraic dependence on $\ell_{\rm max}$, in contrast to $\langle n\rangle$ which depends only on the lower cutoff $\ell_0$.

The loop statistics of protein-DNA clusters are not easily accessible in experiments. Instead, the most relevant results for which this model can provide
 insight come from ChIP-Seq experiments. These experiments yield data for the enrichment of bound ParB as a function of genomic position on the DNA, providing
a measure of the average protein binding profile of ParB on DNA~\cite{Breier,Sanchez}.
 In the LC model, the ParB density profile along DNA can be calculated from:
\begin{eqnarray}
\label{Ptot}
P_{LC}(s)&= &\frac{1}{Z_{LC}}\sum_{n=0}^{m-1} \frac{(m-1)!}{(m-n-1)!n!}\exp(-n J_S)   \\
&&\int_{\ell_0}^{\infty} d \ell_1\ell_1^{-d \nu} ...\int_{\ell_0}^{\infty}d \ell_n \ell_n^{-d \nu} P_{n}\left (s, \{\ell_i\}\right) \nonumber 
\end{eqnarray}
where $Z_{LC}$ is given in Eq.~\eqref{eq:Z}. Here,  $P_{n}\left (s, \{\ell_i\}\right)$ represents the multiloop ParB binding profile with $n$ loops of length $\{\ell_i\}=\{\ell_1,...,\ell_n\}$.
For simplicity, we approximate this multiloop profile by the analytical 1-loop conditional probability, $P_1(s,\ell | \overline{\text{loop$(s)$}})$, with the loop length
 equal to the accumulated loop length, i.e. $\ell\rightarrow \sum_i \ell_i$,
weighted by the loop probability
 $p_{\rm loop}(s, \{\ell_i\})\approx\sum_{i=1}^n\ell_i\rho(s,m,\ell_i,\ell)$. In the expression for the loop probability,
 $\rho(s,m,\ell_i,\ell)$ is defined as the contribution to the loop density of a loop of length $\ell_i$ in a cluster
 of $m$ proteins with a total accumulated loop length $\ell$, and we neglected correlations between contributions from different loops.
 Furthermore, we approximate  $\rho(s,m,\ell_i,\ell)$ by using a generalization of the 1-loop expression in Eq.~\eqref{eq:4},
 \begin{widetext}
\begin{equation}
\rho(s,m,\ell_i,\ell)
\approx \frac{\min(s,\ell_i,m+\ell-s)}{\Theta(m+\ell-2\ell_i)\left[\ell_i^2+\ell_i(m+\ell-2\ell_i)\right]+\Theta(2\ell_i-(\ell+m))
\left(\frac{m+\ell}{2}\right)^2}\,.
\label{loopdensity}
\end{equation}
\end{widetext}
In the analysis above, we aimed to capture the effects of multiple loops in a simple way by assuming statistical independence of the loops, and by using
 the analytical 1-loop expressions to approximate the impact of loop formation on the loop density and the ParB binding profile of the protein-DNA complex. To test the validity of these approximations, we performed MC simulations of the complete LC model. We find that the numerically obtained
average loop probability is in reasonable agreement with our approximate expression for the multiloop density, as shown in Figure~\ref{fig4}c. Thus, despite
 the simplicity of our approach, the analytical model provided here captures the essential features of looping in protein-DNA clusters.

The protein binding profile $P_{\rm LC}(s)$ around a {\it parS} site is calculated by averaging the static binding profile for different
  total loop numbers and loop lengths using the  Boltzmann factor (see Eq.~(\ref{eq:Z}))
 from the Looping and Clustering model as the appropriate weighting factor. The resulting
 expression in Eq.~\eqref{Ptot} for the protein binding profile of a protein-DNA cluster is the central result of this paper. We use this expression to compute binding profiles for the full
 Looping and Clustering model, which are shown in Figure~\ref{fig6} as a function of  the distance $s$ to {\it parS} for $m=100$, 200, and $400$.
By construction, the site $s=0$ corresponding to {\it parS} is always occupied, and thus $P(s=0)=1$ for all values of the spreading energy $J_S$.
 This feature of the LC model captures the assumed strong affinity of ParB for a {\it parS} binding site.
 For $J_S=4$, the binding profile  converges to a  triangular profile,
 implying a very tight cluster of proteins on the DNA with almost no loops. The triangular profile in this case results from all the distinct configurations
 in which this tight cluster can bind to DNA such that one of the proteins in the cluster is bound to {\it parS}, and therefore the probability drops linearly
 to 0 at $s\approx m$. The same triangular binding profile was observed for the S\&B model in the strong coupling limit $J_S\to\infty$~\cite{Broedersz}.
 Interestingly, as $J_S$ becomes weaker, we observe a faster decrease of the binding profile near {\it parS} together with a broadening of the tail
 of the distribution for distances far from {\it parS}. This behavior results from the increase of the number of loops that extrude from the
 ParB-DNA cluster with decreasing spreading bond strength $J_S$. The insertion of loops in the cluster allows binding of ParB to occur
 at larger distances from {\it parS}. Thus, the genomic range of the ParB binding profiles is set by  $s_{\rm max}\approx m+\langle \ell \rangle$,
 where the average cumulated loop length $\langle \ell \rangle$ is controlled by $J_S$ (see Eq.~\eqref{looplength}) and $m$. These results illustrate how the full average binding profile is controlled by the spreading bond strength $J_S$: the weaker $J_S$,
 the looser the protein-DNA cluster becomes, which results in a much wider binding profile of proteins around {\it parS}. In the limit $J_S\rightarrow0$, the LC model quantitatively reduces to the statistics of non-interacting loops, as shown in the inset of Figure~\ref{fig6}. In this case, the binding profiles exhibit asymptotic behaviour $P_{\rm LC}(s)\propto s^{-d \nu}$ for large $s$, as in the Stochastic Binding model~\cite{Sanchez}.
Interestingly, we observe a weaker scaling $P_{\rm LC}(s)$ with $s$ at intermediate genomic distances, which we attribute to the reduced loop density near {\it parS} (see Figure~\ref{fig4}c).

\begin{figure}
\centering{
\includegraphics[width=0.7\linewidth]{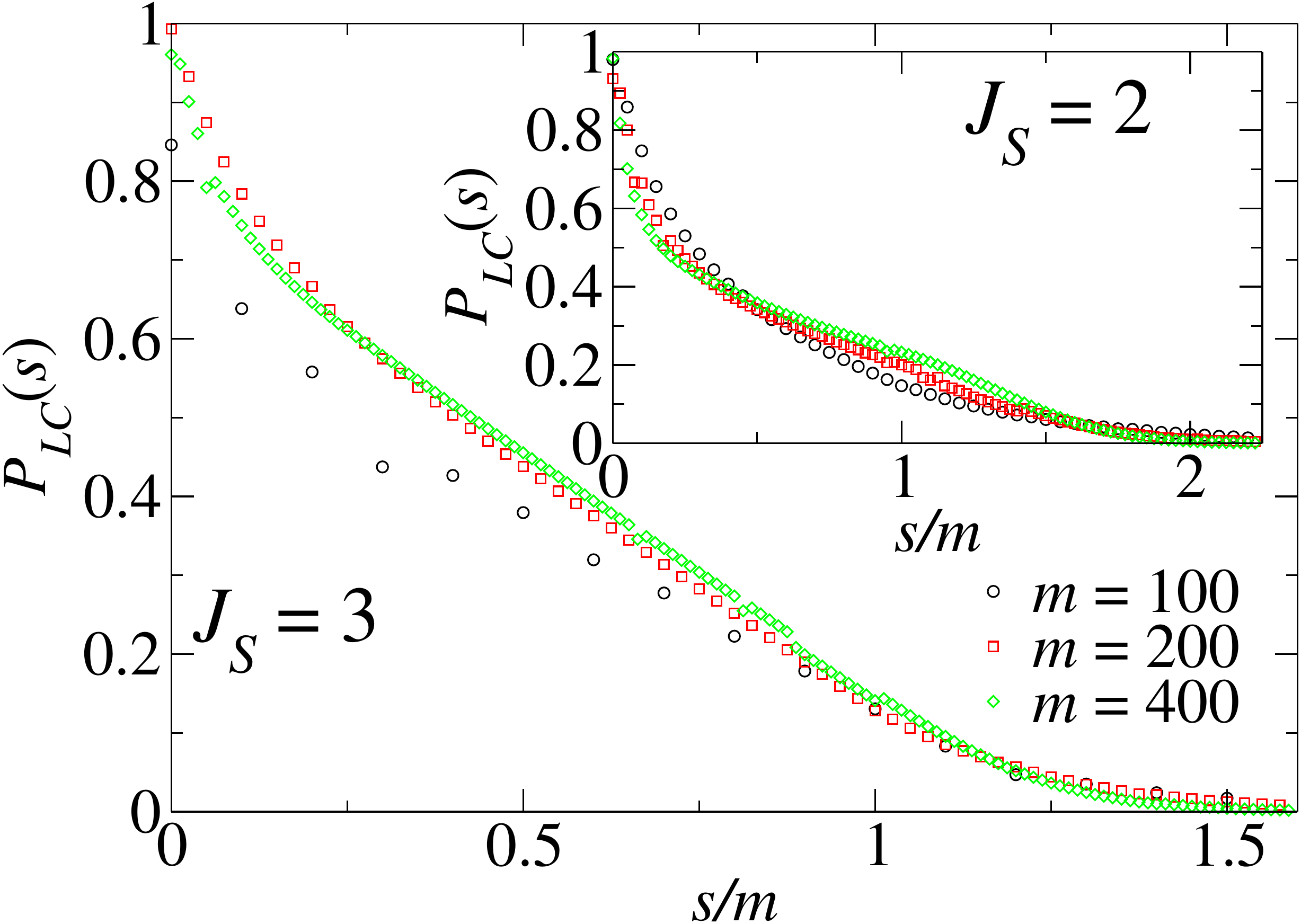}
}
\caption{\label{J=0} Scaling function of the ParB binding profile for different total protein numbers $m$ (same data as Figure~\ref{fig6}).
 The data for different total protein numbers $m$ are plotted versus the dimensionless genomic distance $s/m$ from {\it parS} (main graph: $J_S=3$, inset: $J_S=2$).}

\end{figure}

To investigate how  the functional shape of the binding profile is determined by the total number of proteins in the cluster,
 we plot the  binding probability versus
 the scaled variable $s/m$ for $m=100$, 200, and $400$, as shown in Figure~\ref{J=0}. For fixed $J_S$, the data approximately collapse onto
 a single curve as a function of the scaled distance  $s/m$. This implies that the functional shape of the ParB binding profile is largely determined by
 the spreading bond strength $J_S$, while the number of proteins in the cluster determines the width of the profile.

\section{Discussion} 

The Looping and Clustering model introduced here allows us to access the average binding profile of proteins making up a large 3D protein-DNA complex.
 In our model, the formation of a coherent cluster of ParB proteins is ensured by a combination of spreading and bridging bonds between DNA bound proteins,
 which together can drive a condensation transition in which all ParB proteins form a large protein-DNA complex localized around a {\it parS} site~\cite{Broedersz}.
 We do not assume, however, that this protein-DNA cluster is compact. Indeed, loops of protein-free DNA may extend from the cluster, which strongly influences
 the average spatial configuration of proteins along the DNA.
 In the LC model, the formation of loops in the protein-DNA cluster is controlled by the strength of spreading bonds, i.e. the bond between proteins bound
 to nearest neighbor sites on the DNA. Specifically, for every protein-free loop of DNA that extends from the cluster, a single spreading bond between
 two proteins within the cluster must
 be broken. Thus, if the spreading interaction energy, $J_S$, is sufficiently small, thermal fluctuations will enable the transient formation and breaking of spreading bonds,
 thereby allowing multiple loops of DNA to emanate from the protein cluster (See Figure~\ref{fig:schematic}).

 Conceptually, the spreading bond interaction determines how ``loose" the protein-DNA cluster is, which directly impacts the ParB binding profiles. When $J_S$ is large,
 loop formation is unlikely, resulting in a compact protein-DNA cluster with a corresponding triangular protein binding profile centered around {\it parS}~\cite{Broedersz}.
 At intermediate $J_S$, the protein-DNA cluster becomes looser with the formation of loops, resulting in a binding profiles that are more strongly peaked
around {\it parS} but with far-reaching tails. Importantly, the LC model enables us to establish a link between the Spreading \& Bridging model and the Stochastic Binding model~\cite{Sanchez}. The first used a microscopic approach based on the types of interactions between proteins on the DNA polymer, while the second employed a more macroscopic approach based on the polymer configurations around a dense sphere of proteins.
In the limit  $J_S\to0$, the LC model is consistent with the Stochastic Binding model with a profile of the form~\cite{Sanchez} given by
 $P(s)\propto s^{-d \nu}$(inset Figure~\ref{fig6}a). Thus, the LC model offers a description for a broad parameter regime, connecting two
 limits investigated in preceding studies~\cite{Broedersz,Sanchez}.

The Looping and Clustering model, which we introduce to calculate the binding profile of ParB-like proteins on the DNA,
is a simple theoretical framework similar to the Poland-Scheraga model for DNA melting~\cite{Poland,Everaers}. An important difference in the LC model with respect to
 the homogeneous  Poland-Scheraga model,
 is that translational symmetry is broken due to the presence of a {\it parS} site at which a protein is bound with a high affinity. Thus, the protein-DNA cluster can adopt a wide range of configurations as long as one of the proteins is bound to the {\it parS} site. As a result, loops are effectively
 excluded in the vicinity of {\it parS}. The central new result of this work is a simple way of computing the protein binding profiles around such a {\it parS} site in terms of molecular interactions parameters. We show that the  binding profiles predicted by this model are sensitive to both the expression level of proteins and the
 spreading interaction strength, which directly controls the formation of loops in the protein-DNA cluster. The LC model predicts a profile in good quantitative agreement with binding profiles measured with ChIP-Seq on the F-plasmid of {\it E. coli}, as shown in Fig.~\ref{fig6}c. Importantly, from this analysis we extract the spreading interaction strength  $J_S\approx1 k_{\rm B} T$ and  the number of proteins in the cluster $m\approx400$.

 Our results also have  implications for experiments that employ fluorescent labelling of DNA loci by exogenous ParB{\it s}~\cite{Chen,Saad}.
 Indeed, our model can be used to investigate how the protein interaction strengths determine the 3D structure and mobility of the ParB-DNA cluster,
 as well as the tendency of multiple ParB foci to adhere to each other.
This model thus provides an insightful quantitative tool that could be employed to analyze and
 interpret ChIP-Seq and fluorescence data of ParB-like proteins on chromosomes and plasmids.\\

 \begin{acknowledgments}
This project was supported by the German Excellence Initiative via
 the program NanoSystems Initiative Munich (NIM) (C.P.B.), the Deutsche Forschungsgemeinschaft (DFG) Grant TRR174 (C.P.B),
and the National Science Foundation Grant PHY-1305525 (N.S.W.). We also thank J.-Y. Bouet for helpful comments on the manuscript.
The authors acknowledge financial support from the \textit{Agence Nationale de la Recherche} (IBM project ANR-14-CE09-0025-01)
 and from the CNRS D\'efi Inphyniti (\textit{Projet Structurant}  2015-2016). This work is also part of the program ``Investissements d’Avenir''
 ANR-10-LABX-0020 and Labex NUMEV (AAP 2013-2-005, 2015-2-055, 2016-1-024).
\end{acknowledgments}

\appendix

\section{Monte Carlo simulations and numerical integration procedures}

\subsection{Monte Carlo procedure}

Using the partition function, we can formulate an effective 1D Hamiltonian
 for the LC model, which explicitly accounts for the balance between spreading bonds and loop entropy:

\begin{equation}
    \mathcal H_{LC}=-J_S\,\sum_{i=1}^{L-1}\,\phi_i\phi_{i+1}+d\nu\sum_{i=1}^{n}\ \ln(\ell_i+\ell_0)\,.
\end{equation}

\begin{figure}
    \centering\includegraphics[width=.75\linewidth]{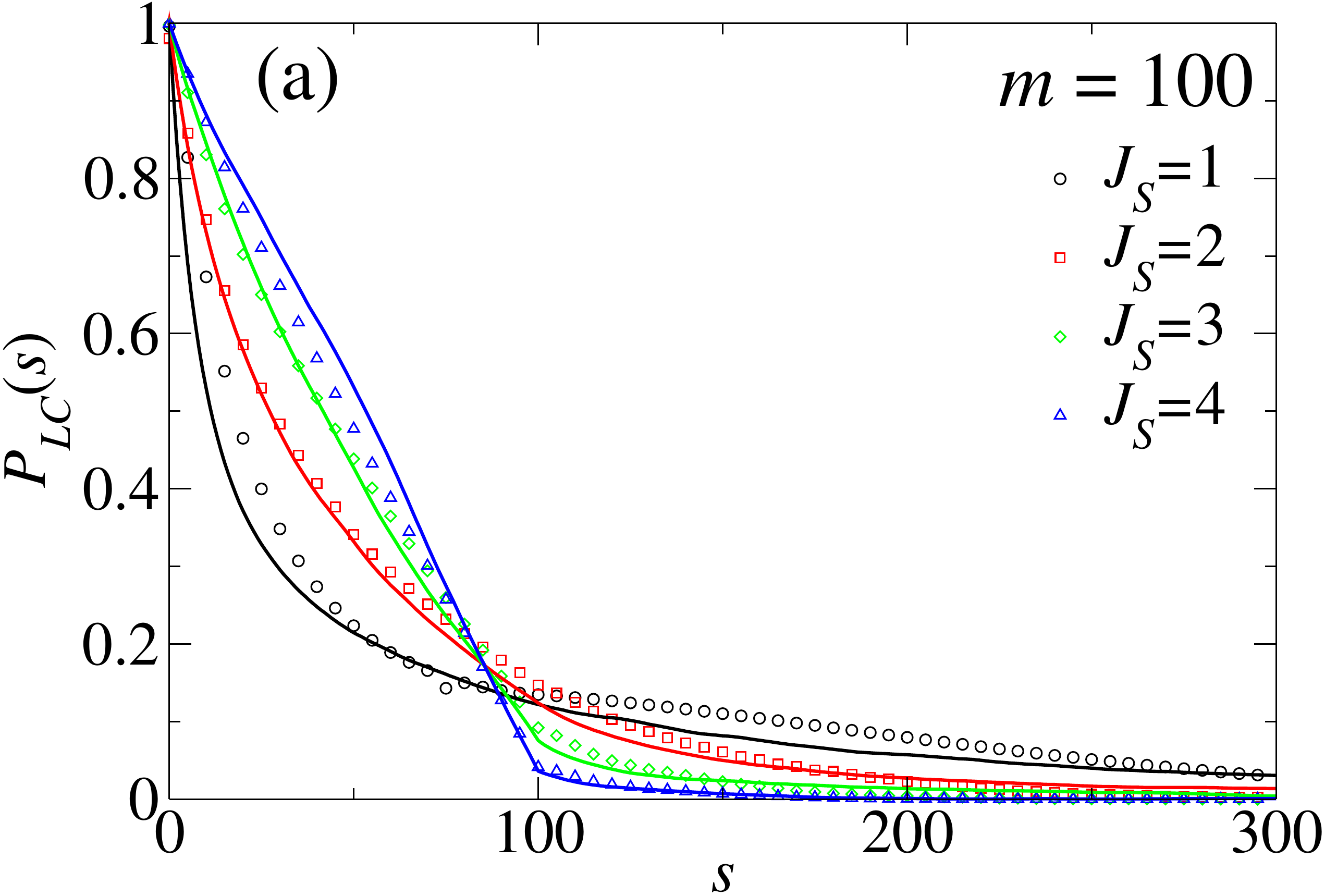}
\caption{\label{fig:MCcomparison}The binding profile obtained with the analytic approach (symbols) are compared to MC simulations (dashed lines)
for $m=100$, $\ell_0=10$, and $J_S=1,2,3$ and 4. }
\end{figure}

This effective Hamiltonian is useful to perform Monte Carlo simulations of the
 model as a benchmark for the approximations performed in the analytical approach (see Fig.\ref{fig:MCcomparison}).
The proteins are modelled as particles that bind/unbind onto sites of a one-dimensional lattice with free boundary conditions.
 The lattice size $L=4000$ is chosen to prevent finite size effect for the range of
 proteins considered. Note that, in these MC simulations, the total size of the loops is limited to $L-m$.

The simulations are performed with the standard Metropolis rules:
\begin{enumerate}
\item Propose a move of a particle randomly chosen to a random empty site of the lattice (conserved order parameter).
A MC iteration step consists of $m$ attempts of move.
\item Calculate the difference of energy $\Delta \mathcal H = \mathcal H_f-\mathcal H_i$ between final and initial configurations.
\item If $\Delta \mathcal H<0$, the move is accepted with probability 1, otherwise it is accepted with
 probability $\exp(-\beta\Delta \mathcal H)$.
\end{enumerate}

The system is set initially with all particles in a single cluster ($J_S=\infty$), and then thermalized to the actual $J_S$ of the
simulations ranging from 1 to 4 (see Fig.~\ref{fig:MCcomparison}). The sampling starts after thermalization of the system
(40000 MC iterations). A sampling of the systems configuration is performed every 100 MC iterations. All MC averages have been
 performed over $10^7$ configurations, $\nu=0.588$, $L=4000$ and $\ell_0=10$.
The numerical results of this Monte Carlo simulation are in good agreement with our approximate analytic results, as shown in Fig.~\ref{fig:MCcomparison}.

\subsection{Numerical integration}

To evaluate the binding profile $P_\text{LC}(s)$, we proceeded as follows. We carried out the evaluation of the simplified expression in Eq.~\eqref{Ptot} using numerical summation and integration. We truncated the summation at $n=15$, instead of going up to $m-1$, based on the corresponding average number of loops of Fig.~\ref{fig6}. Finally, we introduced an upper cutoff for the loop-length, $\ell_{\rm max}=100$, instead of going up to infinity. We confirmed that shape of the binding profiles does not change significantly for higher values of $\ell_{\rm max}=100$.

The numerical evaluation of the multidimensional integrals in Eq.~\eqref{Ptot} have been performed with an accuracy and precision of respectively 2 and 3 effective digits in the final results. We have carried out convergence tests of the curve shapes in order to assess our parameters choice and rule out numerical instabilities. All computations have been performed by routines written in the Wolfram Language and executed by the Mathematica software suite (version 10 and 11).

\section{Formal connection between the LC model and a Lattice Gas with renormalized coupling}
For $m$ and $n \gg 1$ (thermodynamic limit), we can formulate a saddle point approximation to evaluate the partition function and $\langle n \rangle$,
by  approximating  the entropic (factorial) term in $Z$ (Eq.(\ref{n0})) using the standard entropy of mixing for placing $n$ loops on  $m-1$ possible sites. This approach gives physical insight into how the loop entropy contributes to a renormalized  protein-protein interaction and how the competition between this renormalized interaction and the entropy of mixing controls $\langle n \rangle$. Taking the thermodynamic limit leads to a partition function:
\be
     Z' =  \int_0^\infty d\rho_{\ell} \, \, \exp \left[ - (m-1) F_{\rm eff}(\rho_{\ell}) \right],
\ee
where  $\rho_{\ell} = n/(m-1)$ is the  concentration of loops ($0 \leq \rho_{\ell} \leq 1$) and
\be
      F_{\rm eff}(\rho_{\ell}) =  \rho_{\ell} J'_S +  \{  [1 - \rho_{\ell}] \ln[1 - \rho_{\ell}] +
   \rho_{\ell} \ln[\rho_{\ell}] \}
\ee
an effective free energy where  $J'_S = J_S + \ln \alpha_0$  is a loop activation energy renormalized by the cost in loop entropy with
 $\alpha_0 = \ell_0^{d \nu - 1} (d \nu - 1)$.  In the limit $m \rightarrow \infty$, the approximate partition function $ Z'$ becomes
 exact and can be evaluated exactly in the saddle point approximation by minimizing $F_{\rm eff}$. The solution, $\rho_{\rm SP}$,
 to the saddle point equation, $dF_{\rm eff}(\rho_{\ell})/d\rho_{\ell} = 0$, is
\be
      \rho_{\rm SP} =
      \frac{1}{1+ e^{J'_S}}.
\ee
The entropic contribution to $F_{\rm eff}$ (second term) vanishes at  $\rho_{\ell} = 0$ and 1, and reaches a minimum at $\rho_{\ell} = 1/2$,
 which is the exact result for $\rho_{\rm SP}$ at vanishing renormalized loop activation energy  $J'_S$  because the entropy of mixing is then maximized.
 For  $J'_S > 0$, $\rho_{\rm SP}$ decreases from 1/2 to vanish in the limit $J'_S \rightarrow \infty$ as $\rho_{\rm SP} \rightarrow e^{-J'_S}$. In this limit only the no  and one loop states contribute and the asymptotic behavior can be simply obtained by a series expansion of the partition sum and the corresponding expression for $\langle n \rangle$.
 The saddle point result leads to
\be
     n_{\rm SP} = \frac{m - 1}{1+ e^{J'_S}},
\ee
which turns out to be the exact result, thanks to compensating errors,  for $\langle n \rangle$ for finite $m$
 (which can be obtained by differentiating the exact $Z$ with respect to $J_S$, see Eq.(\ref{n1})). For example, for $\ell_0 = 10$, $d = 3$, and $\nu = 0.588$, $\alpha_0 = 4.437$ and $\ln(\alpha_0) = 1.49$, which is not negligible if $J_S=\mathcal O(1)$.

\end{document}